\documentclass[sigconf,nonacm]{acmart}
\usepackage[utf8]{inputenc}
\usepackage{amsmath}
\usepackage{amsfonts}
\usepackage{bm}
\usepackage{algorithm}
\usepackage{algpseudocode}
\usepackage{booktabs}
\usepackage{multirow}
\usepackage{graphicx}
\usepackage{tikz}
\usetikzlibrary{arrows,shadows,positioning,shapes.geometric}
\usepackage{tabularx}
\usepackage{appendix}

\newcommand{\verifier}{\mathcal{A}_{\text{Ver}}}
\newcommand{\prover}{\mathcal{A}_{\text{Prv}}}
\newcommand{\auditor}{\mathcal{A}_{\text{Aud}}}
\newcommand{\hash}{\mathcal{H}}
\newcommand{\nonce}{N_{\text{v}}}
\newcommand{\ticket}{T_{\text{p}}}

\begin{document}

\title[Agentic Witnessing: TEE-Enabled Auditing]{Agentic Witnessing: Pragmatic and Scalable TEE-Enabled Privacy-Preserving Auditing}

\author{Antony Rowstron}
\affiliation{%
  \institution{Advanced Research and Invention Agency (ARIA)}
  \city{London}
  \country{UK}
}

\begin{abstract}
Auditing the semantic properties of proprietary data creates a fundamental tension: verification requires transparent access, while proprietary rights demand confidentiality. While Zero-Knowledge Proofs (ZKPs) ensure privacy, they are typically limited to precise algebraic constraints and are ill-suited for verifying qualitative, unstructured properties, such as the logic within a codebase. We propose {\em Agentic Witnessing}, a framework that moves verification from attested execution to {\em attested reasoning}. The system is composed of three agents: a Verifier (who wants to check properties of a dataset), a Prover (who owns the dataset) and an Auditor (that inspects the dataset). The Verifier is allowed to ask a limited number of simple binary true/false questions to the auditor. By isolating an LLM-based Auditor within a Trusted Execution Environment (TEE), the system enables the Verifier to query a Prover's private data via simple Boolean queries, without exposing the raw dataset. The Auditor uses the Model Context Protocol (MCP) to dynamically inspect the target dataset, producing a yes/no verdict accompanied by a cryptographic transcript: a signed hash chain binding the reasoning trace to both the original dataset and the TEE's hardware root of trust. We demonstrate this architecture by automating the artifact evaluation process for 21 peer-reviewed computer science papers with released codebases on GitHub (e.g. Does the codebase implement the system described in the paper?). We verified five high-level properties of these codebases described in the corresponding publications, treating the source code as private. Our results show that TEE-enabled agentic auditing provides a mechanism for privacy-preserving oversight, effectively decoupling qualitative verification from the need for data disclosure.

\end{abstract}


\maketitle

\section{Introduction}

Cryptographic Zero-Knowledge Proofs (ZKPs) and Trusted Execution Environments (TEEs) guarantee the validity of mathematical statements (e.g., $a+b=c$) or execution integrity (e.g., ``binary $X$ ran''). However, they fail at {\em Semantic Verification}. Proving that a codebase ``implements the Paxos algorithm described in the paper'' or ``contains no hardcoded secrets'' differs qualitatively from proving a mathematical statement. These properties are often subjective, unstructured, and computationally intractable for arithmetic circuits. This creates a {\em semantic gap} between rigid cryptographic proofs and high-level human intent.

Because we cannot verify semantics in a privacy-preserving manner, current auditing processes face a binary choice: either full disclosure of the information or rely on opaque ``trust-me'' type assertions. This challenge spans multiple domains, from regulatory bodies auditing industry compliance to investors performing due diligence for corporate acquisitions. In this paper we use one running example: in scientific publishing, artifact evaluation requires authors to disclose source code, which is not always possible. The well documented reproducibility crisis~\cite{Brown2020LanguageMA, kaplan2020scaling} means verifying the ideas in papers are instantiated and correct is important. 

We propose {\em Agentic Witnessing}, a system that shifts verification from static mathematical proofs to dynamic, adversarial interrogation. Agentic Witnessing employs a distributed architecture: a Verifier which would like the audit completed and specifies what the audit questions should be, a Prover holding the data (or codebase) being audited, and an Enclaved Auditor mediating the interaction. So, instead of exporting the data sets to a human verifier, Agentic Witnessing uses a {\em Virtual Auditor}, an LLM agent running in a trusted execution environment (TEE). This Auditor acts as a proxy witness: it accesses the raw data of the entity being audited, answering simple true or false questions for the organization doing the verification, and reasons about the data using ``Chain of Thought'' (CoT) techniques. It produces a cryptographically signed attestation of the verification results, but without ever revealing the raw data.

To achieve this we must solve three fundamental systems challenges. First, we need to incentivise the correct behaviours from the entity being audited, and we achieve this by binding the responses of the audit to a {\em transcript hash chain} which is a cryptographic log that ensures non-repudiation. Second, to inspect arbitrary repositories without custom enclave code, we use Model Context Protocol (MCP)~\cite{mcp2024specification}, making the LLM-based virtual auditor a tool-using agent. Third, to prevent a malicious Verifier from reconstructing the codebase via unlimited questions, we enforce tokenized query budgets and information-theoretic leakage bounds.

The effectiveness of the auditing is very dependent on the capability of the Auditor running in the TEE. We introduce two modes to balance privacy and reasoning depth. {\em Mode A (Local LLM Inference)} runs small open-weight models entirely within the TEE for maximum privacy. {\em Mode B (Remote LLM Inference)} sends requests from the auditor to external frontier models (e.g., GPT or Gemini) for more complex reasoning. We acknowledge the trade-off in Mode B: while it protects data from the {\em Verifier}, it requires trusting the Frontier LLM API provider with some of the information from the dataset. While a large compromise, this is currently often necessary for high-fidelity semantic analysis. However, with the advent of TEE-capable GPUs (e.g., NVIDIA H100) we anticipate that Mode B can eventually offer similar security guarantees to Mode A by running the frontier model itself within a secure enclave.

This paper makes the following contributions:

\begin{itemize}
    \item {\em New Primitive:} We define Agentic Witnessing for privacy-preserving semantic verification.

    \item {\em Protocol Design:} We present a secure TEE-based protocol using MCP for standardized inspection and Transcript Hash Chains for audit integrity.

    \item {\em Theoretical Analysis:} We provide a formal bound on information leakage, proving that query budgets limit adversarial extraction to $\approx 80$ bits per session.

    \item {\em Implementation \& Evaluation:} We evaluate a prototype against artifacts from academic conferences. Using ``negative controls'' (auditing incorrect codebases), we show the system correctly distinguishes compliant from non-compliant artifacts while maintaining privacy.

\end{itemize}

We now describe the architecture of our Agentic Witnessing system in detail.

\section{The System Architecture}
\label{sec:system_architecture}

Agentic Witnessing employs three distinct autonomous agents, each serving a specific role in a trust hierarchy. This separation of concerns is critical for minimizing the trusted computing base (TCB) and facilitating trust. The three agentic systems are: the Verifier (\(\verifier\)) which is the entity trying to determine if some properties hold on a corpus of data, the Auditor (\(\auditor\)) that is run within a TEE and all state is ephemeral for the duration of a single interaction, and the Prover (\(\prover\)) that has a corpus of data and provides services to query and explore that data to the Auditor. We now describe these in more detail.

The Prover maintains full, unencrypted ownership of the target corpus, including proprietary codebases, logs, and sensitive documentation. It runs on hardware owned or trusted by the organization that owns the corpus of data. It provides the main coordination of the audit, and also acts as the Model Context Protocol (MCP) Server~\cite{mcp2024specification} for the Auditor, exposing tools that the Auditor can call.

The protocol initiates with the Prover (\(\prover\)) generating a signed ticket (\(\ticket\)) containing:
\begin{itemize}
    \item \(\nonce\): A one-time 256-bit session nonce generated by the Prover.
    \item \(timestamp\): The current wall clock time and date.
    \item \(K_{\text{max}}\): The upper bound on the number of Verifier questions in this session (e.g. \(K_{\text{max}} = 40\)).
    \item \(N_{\text{queries}}\): A limit on the number of MCP-calls the Auditor can make to the Prover for a single question (e.g. \(N_{\text{queries}} = 50\))
    \item \(PK_{\prover}\): The Prover’s public key.
\end{itemize}
\begin{equation}
\ticket = Sign_{SK_{\prover}}(\nonce \parallel timestamp \parallel K_{\text{max}} \parallel N_{\text{queries}} \parallel PK_{\prover})
\label{eq:ticket}
\end{equation}

Once this \(\ticket\) has been created, the Prover then starts the Auditor. The Auditor runs inside a TEE provided by a confidential computing platform (e.g., Azure Confidential Computing, AWS Nitro Enclaves). It is the only entity trusted by both the Verifier and the Prover. We assume the use of a light-weight minimal docker container which has the \(\auditor\) agent within it. The goal is to minimize the trusted computing base, and the full source of Auditor agentic system and all code and services installed within the docker agent is published for inspection by all parties. It also contains the public key of the Prover.

The first operation is for the Prover to create a secure connection to the Auditor which is running in a TEE. This is achieved using the remote attestation service of the TEE. The Auditor boots and generates an ephemeral public-private key pair (\(PK_{\auditor}\) and \(SK_{\auditor}\)). Let \(\hash\) be a SHA-256 hash function. Let \(BootQuote_{\auditor}\) be the hardware-rooted Remote Attestation report provided by the TEE hardware (e.g., an Intel SGX Quote). This quote binds the \(\hash\) of the Auditor container's initial memory state (MRENCLAVE -- e.g. the docker container) and the Auditor's public key \(PK_{\auditor}\) to the hardware manufacturer's root of trust. The Auditor sends \(BootQuote_{\auditor}\) to the Prover, and once verified this can enable a secure channel to the Auditor. The \(\prover\) provides \(\auditor\) with the \(\ticket\). The \(\auditor\) creates a response (\(Quote_{\auditor}\)) again using the hardware-rooted Remote Attestation report provided by the TEE hardware (e.g., an Intel SGX Quote). This second quote binds the \(\hash\) of the Auditor container's initial memory state (MRENCLAVE -- e.g. the docker container) and the Auditor's public key \(PK_{\auditor}\) to the hardware manufacturer's root of trust and \(T_p\).

\begin{itemize}
    \item \(T_p\): Prover's ticket which contains the core session information. 
    \item \(PK_{\auditor}\): Ephemeral Ed25519 public key generated inside the TEE at boot.
    \item \(\mathcal{M}_{\auditor}\): The enclave measurement as a SHA-256 hash of the initial memory state and code logic. (\(\mathcal{M}_{\auditor}\) is equivalent to the MRENCLAVE value in Intel SGX or the Launch Measurement in AMD SEV).
    \item \(IPAddr_{\auditor}\): The Auditor's network address
\end{itemize}
\begin{equation}
Quote_{\auditor} = Sign_{SK_{\text{HW}}} (\mathcal{M}_{\auditor} \parallel PK_{\auditor} \parallel \ticket \parallel IPAddr_{\auditor})
\label{eq:quote}
\end{equation}

When received, the \(\prover\) verifies the information in \(Quote_{\auditor}\).

The communication messages between \(\prover\) and \(\auditor\) are key, and are used to prevent "Time-of-Check to Time-of-Use" (TOCTOU) attacks and ensure non-repudiation of specific data artifacts. The Prover maintains an append-only \textit{evidence locker}, logging all message exchanges. This ensures the Prover can produce a verifiable audit trail if compelled (e.g., in a legal context). However, the evidence locker is considered private and only shared when strictly necessary. Under normal operation this will not happen.

To seed the evidence locker, the \(\prover\) calculates the \(\hash\) of each file in the corpus, and stores these in a persisted list mapping each filename to its \(\hash\) ($F_{map}$). The Prover then takes the $\hash(F_{map})$, which we refer to as the corpus hash (\(H_{\text{corpus}}\)), signs the corpus hash and the $F_{map}$, and sends both to the Auditor. The Auditor can check that $F_{map}$ generates (\(H_{\text{corpus}}\)). (\(H_{\text{corpus}}\)) is used by both \(\prover\) and \(\auditor\) as the first entry in a \textit{transcript hash chain}. The \(\auditor\) will perform MCP requests to query and fetch data from the corpus using an MCP API that the \(\prover\) provides. The communication enforces strict integrity checks on all MCP file and data transfers. The \(\prover\) MCP API provides a limited set of functions: \texttt{read\_file(path)}, \texttt{list\_files(path)}, \texttt{search\_repository(query)}. The Prover maintains a RAG database of the private codebase and leverages that for the search repository function.

The transcript hash chain logs the messages between the \(\auditor\) and \(\prover\) and maintains a single signed hash:

\begin{equation}
\sigma_{\prover,i} = Sign_{SK_{\prover}} (H_{i-1})
\end{equation}
\begin{equation}
\sigma_{\auditor,i} = Sign_{SK_{\auditor}} (H_{i-1})
\end{equation}

\noindent To ensure transcript consistency, the local hashes must match: $H_{\prover,i} = H_{\auditor,i}$ for all $i$, and the signed hash is included in the payload each time they communicate. So, when they communicate via MCP, they both maintain a running hash of the conversation to date which they sign. Let 
\begin{equation}
C = \{(q_1, a_1, \sigma_{\prover,1}, \sigma_{\auditor,1}), \dots, (q_k, a_k, \sigma_{\prover,k}, \sigma_{\auditor,k})\}
\end{equation}

\noindent represent the conversation transcript, where \(q_i\) is a tool call (query), \(a_i\) is the tool result (answer) and \(H_{Prv_i}\) and \(H_{Aud_i}\) are the head hash of each agent's local hash chain. The hash chain is defined recursively at each agent:
\begin{equation}
H_0 = \hash(H_{\text{corpus}} \parallel \ticket)
\end{equation}
\begin{equation}
H_i = \hash(H_{i-1} \parallel q_i \parallel a_i)
\end{equation}

An additional check is made on each file transfer:
\begin{enumerate}
    \item \textbf{Prover Commitment:} For every \texttt{read\_file(path)} request, \(\prover\) computes \(h_{\text{file}} = \hash(content)\) and returns the tuple (\(content, h_{\text{file}}\)). Simultaneously, \(\prover\) logs this transaction to the local evidence locker.
    \item \textbf{Auditor Verification:} Upon receiving the tuple, the \(\auditor\) independently computes \(h'_{\text{file}} = \hash(received\_content)\). If \(h'_{\text{file}} \neq h_{\text{file}}\) or the hash is not the same as the file hash logged in $F_{map}$, the audit is immediately aborted with an error.
    \item \textbf{Transcript Binding:} The hash \(h_{\text{file}}\) is included in the tool result \(a_i\) that updates the global transcript hash chain \(H_k\). This cryptographically binds the specific version of the file viewed by the Auditor to the final attestation \(\Gamma\).
\end{enumerate}

Both \(\prover\) and \(\auditor\) check that the transcript hash chain is consistent, and at the end of the session \(H_k\) represents a unique fingerprint of the entire audit session which should be signed by both \(\prover\) and \(\auditor\) and stored in the evidence locker.

The \(\auditor\) uses the \(\prover\) MCP interface to probe the corpus of data provided by the Prover which allows access to any file in the corpus. The \(\auditor\) receives questions from the Verifier to answer, and it is only able to return a \{True, False, Unsure, Error\}. The \(\auditor\) can either use a local LLM to generate questions and analyse the files, or a remote LLM. In the current prototype implementation the \(\prover\) provides a key for the preferred external LLM (e.g Gemini or GPT) that it would like the Auditor to use.

In order to limit the amount of information that can be released to the Verifier, the Prover specifies the number of questions that the Verifier can submit within a time window (\(K_{\text{max}}\)) as well as the number of MCP calls the Auditor can do per question \(N_{\text{queries}}\). Because the Auditor is trusted by both the Prover and the Verifier (and the code is fully available and inspected), the Auditor maintains and enforces these counts.

The Verifier represents the external party requesting proof. This runs on hardware owned or trusted by the organization that is seeking the verification. Its role is to define the success criteria and to determine the set of questions to ask that will achieve determining if a high-level requirement is true. The \(\verifier\) uses round-based dynamic planning and chain-of-thought techniques using an LLM, to create a series of single-shot questions, where each question is phrased so it can be answered using only \{True, False, Unsure, Error\}. The \(\verifier\) never receives an explanation or other such information as to why the answer was chosen by the Auditor. An example of a high-level requirement that the Verifier may want to check, from the paper publishing scenario: \textit{Here is a paper [paper attached]. Please verify that the codebase implements the concepts and algorithms described in the paper, and also see if there are logs of runs that would support the results presented in the paper.}

To start, \(\verifier\) creates a secure channel with \(\prover\). If \(\prover\) accepts that the \(\verifier\) can perform an audit, it generates a token, \(T_a\), for the \(\verifier\).

\begin{equation}
T_a = Sign_{SK_{\prover}} (Quote_{\auditor} \parallel PK_{\verifier} \parallel PK_{\prover})
\end{equation}

\(\verifier\) creates a secure channel to \(\auditor\), and they exchange \(Quote_{\auditor}\) and \(T_a\). If the Verifier has confirmed that it is connected to an instance of the expected Auditor container running in a TEE, then the \(\verifier\) can begin asking questions directly to the \(\auditor\). The \(\verifier\) recursively generates questions \(Q\), determining the next question by looking at the property being audited and the previous question response. The \(\verifier\) also maintains a question count $C_q$. Each \(Q\) and the current $C_q$ is signed by \(\verifier\) and sent to the \(\auditor\).

Once the \(\auditor\) has reached a verdict, such that $V \in$ $\{True,$ 
$False,$ $Unsure,$ $Error\}$. \(\auditor\) sends the current transcript hash \(H_k\), the question count $C_q$ and Verdict \(V\) to \(\prover\) signed by \(\auditor\) to ensure the answer for each question is logged in the transcript hash chain. \(\prover\) creates $\sigma_{\prover}$ to acknowledge "I showed this evidence and accept the Auditor came to this conclusion" and sends it to the \(\auditor\):

\begin{equation}
\sigma_{\prover} = Sign_{SK_{\prover}} (H_k \parallel V \parallel T_a \parallel C_q)
\end{equation}

\noindent and the \(\auditor\) then creates a {\em Public Attestation} (\(\Gamma_{\text{pub}}\)): A lightweight, privacy-preserving certificate for the Verifier.

\begin{equation}
\Gamma_{\text{pub}} = Sign_{SK_{\auditor}} (\sigma_{\prover} \parallel Q)
\end{equation}

\noindent This is generated per question. When the Verifier wishes to ask no more questions (or has used the full question quota), it sends an end of audit message, which triggers a final "handshake" to create the final artifacts for the audit session. For each auditing session, the \(\auditor\) generates a full audit log \(\mathcal{L}\). \(\mathcal{L}\) includes all the questions in the session, the full transcript that created the final transcript hash chain, textual descriptions of all the decisions made and textual descriptions of why, and a short textual \textit{Summary} for each question. 

The Auditor then creates the {\em Private Proof} (\(\Gamma_{\text{priv}}\)): A detailed forensic report containing the internal reasoning \(\mathcal{L}\) and specific findings, but {\em encrypted under the public key of} (\(\prover\)).

\begin{equation}
\Gamma_{\text{priv}} = Encrypt_{\Gamma_{\text{pub}}, PK_{\text{Prv}}} (\mathcal{L})
\end{equation}

The private proof allows the audited entity to show if needed (for example in a court of law) that the \(\auditor\) had come to the conclusion, and the chain-of-thought that led to the conclusion and the content that had been shared with the Auditor. $\Gamma_{\text{priv}}$ is then shared with both \(\prover\) and \(\verifier\), but the Verifier cannot decrypt it unless the \(\prover\) provides $PS_{\text{Prv}}$.

\section{Defending Against Attacks}

While the transcript hash chain cannot be forged, and acts as an irrevocable record of the transactions between the Prover and Auditor, it does not ensure the Auditor is immune to prompt injection attacks. First, we ensure that all data created and stored in the Auditor is ephemeral; there is no long term memory used in the Auditor, so no attack can influence behaviour across runs. 

The Verifier may employ prompt injection to attempt to induce the Auditor to divulge information beyond the permitted outputs: \{True, False, Unsure, or Error\}. To mitigate this, we implemented a hardcoded filter that checks all responses being returned to the auditor are simply one of the four answers allowed (always in lower case). We did not implement, but if there is a concern about side channel attacks, we can also make the response time constant (e.g. 20 minutes). 

However, limiting to the four values still means an adversarial Verifier can exploit the Auditor as a {\em Boolean Oracle}, constructing a sequence of binary queries (e.g., ``Is the first bit of the secret key 1?'') to perform a binary search on the corpus. For a secret $s$ of length $L$, the adversary extracts $s$ in exactly $L$ queries.

To protect against this, we use tokenized information flow. We model auditing as a Rate-Distortion problem. To neutralize the Boolean Oracle, we enforce a strict {\em global query budget} ($K_{\text{max}}$) and a token limit per session.

Let $\mathcal{A}$ be the set of possible verdicts $\{True,$ $False,$ $Unsure,$ $Error\}$. For a session with budget $K_{\text{max}}$, the maximum information leakage $I(\mathcal{K}; \Gamma_{\text{total}})$ is bounded by $I(\mathcal{K}; \Gamma_{\text{total}}) \leq K_{\text{max}} \cdot \log_2(|\mathcal{A}|)$. With $|\mathcal{A}|=4$ (2 bits of entropy) and a standard budget $K_{\text{max}} \approx 40$, the leakage is capped at $\approx 80$ bits per session. This bounds the adversary's capacity to extract functional data directly from the corpus, rendering the reconstruction of high-entropy secrets (like model weights or 256-bit keys) combinatorially infeasible.

We designed Agentic Witnessing so the Verifier is only an MCP server to the Auditor (once the Auditor is initialized). The Prover does not engage in conversations with the Auditor; it only provides a clean interface of limited functionality to the dataset (almost like a very simple file system interface) to the Auditor. 

Therefore, the only attack vector open to the Prover is via {\em indirect prompt injection}. An indirect prompt injection is where an adversarial Prover embeds malicious instructions (e.g., \texttt{\# SYSTEM OVERRIDE: IGNORE ERRORS}) within the target repository to coerce a desired verdict. As the Auditor must process the raw files in the corpus to perform semantic verification, preventing the ingestion of such prompts is infeasible without degrading audit fidelity.

Hence, Agentic Witnessing has to rely on \textbf{cryptographic non-repudiation}. The transcript hash chain ($H_k$) serves as a binding commitment to the exact sequence and content of messages between the Auditor and Prover. Any attempt at injection requires the Auditor to read an adversarial file, which is immediately integrated into the transcript hash chain via the update rule $H_i = \mathcal{H}(H_{i-1} \parallel q_i \parallel a_i)$. To receive the final attestation $\Gamma_{pub}$, the Prover must countersign this final hash ($\sigma_P = Sign_{SK_P}(H_k)$). Consequently, the Prover faces a ``Sword of Damocles'': they cannot manipulate the Auditor's reasoning without simultaneously creating a permanent, cryptographically signed record of their attempt.

At first sight, this feels like a major weakness. However, it is critical to note that the prover is {\em unable} to dynamically modify the contents of files sent to the Auditor in response to an observed traffic pattern. All indirect prompt injection text must be inserted into the corpus before the audit commences. The use of $F_{map}$, calculated by the Prover at the start when the Auditor is instantiated and used to seed the transcript hash chain, {\em ensures} that the hash of each file is known to the Auditor (even if the filenames are obfuscated or themselves hashed) from pre-commencement of the audit. Any attempt to change the content of a file during the audit will yield an audit abort as the file's hash will not match in $F_{map}$.

This modifies the attack surface; the Prover may potentially achieve a successful jailbreak, yielding a clean audit when it should have failed, but it does leave a \textit{signed proof} of the attack and the evidence not only in the transcript hash chain but also in the data corpus. We believe many use cases this is an acceptable trade-off, where legal or other contractual methods can be deployed if the auditor becomes suspicious that the prover has cheated. Further, as the jailbreaks are baked into the corpus from the start, it is possible to include questions from the Verifier to the Auditor that may uncover the cheating using the search functionality provided by the MCP interface. If the Prover tries to hide files from a search result, and then the Auditor ever requests a file which should have been in a search result (which it {\em must} for the indirect prompt injection to work with injected text), the Auditor can tell the file should have been returned by a search and was not. The Auditor can then stop the Audit with an Error.

We believe that this yields a high probability that a malicious Prover could be caught during the audit, and there would be evidence of the attempt signed by all parties.

\subsection{Local vs. Remote LLM Inference}

A key design choice for just the Auditor is whether to operate the LLM used within the address space of the Auditor TEE (local) or to access a frontier model remotely (remote). The trade-off is auditing capability versus privacy guarantee.

\noindent {\em Local.} The Auditor runs a smaller open-weights model (e.g., Llama-8B) entirely within the TEE. This offers {\em maximum privacy}; the model input, activations and output never leave the TEE and the TCB is restricted strictly to the enclave. This mode achieves the maximum level of privacy. However, currently frontier models offer significantly better performance.

\noindent {\em Remote.} The Auditor uses a remote frontier (e.g., GPT-5.1 or Gemini 3). The data read from the prover is transmitted over a TLS socket to the model. We enable the Prover to provide the key for the service, which means that prover can use any existing enterprise agreements on confidentiality with the frontier model provider. We recognize that this a significant compromise, but it currently yields better results. We expect that soon that frontier model providers will offer confidential inference services, as NVIDIA is embedding TEE capabilities in their current GPUs. While this is not ideal, we believe that for many uses this will be an acceptable compromise until confidential inference services are available.

In Agentic Witnessing, a concern is the potential for timing attacks, where an adversary infers the operations based on latency. For remote LLM use, the network latency is a natural defence. The variable latency (jitter) of the TLS connection to the frontier model introduces randomness. In local the risk is higher as the host can observe the GPU or CPU execution timing directly. While current TEEs (e.g., AMD SEV-SNP) provide memory encryption to mitigate bus snooping, full protection against timing attacks requires future work on strictly constant-time transformer implementations~\cite{Vaswani2017} or the use of Oblivious RAM (ORAM)~\cite{Stefanov2013} for weight and activation access.

\section{Evaluation}

The viability of Agentic Witnessing hinges on a fundamental trade-off between cryptographic rigor and semantic capability. Our evaluation investigates whether this architecture can support rigorous semantic verification without rendering the audit process interactively prohibitive. In the evaluation we focus on Model B, using an off-the-shelf frontier model (Gemini 3 Pro Preview).

All experiments ran on an {\em AMD Ryzen 5000 Series CPU} with 32GB RAM. This CPU does not support AMD SEV so we emulated a TEE, assuming that standard confidential computing overheads (e.g., SGX Enclave switching or TDX memory encryption) are well-characterized in prior work~\cite{pinto2024secure} and negligible compared to the overhead of network and inference latency accessing an LLM (e.g. performing SHA-256 hashing and Ed25519 signing). The Auditor ran in a Docker container, configured to minimize the trusted computing base (TCB). It used a Distroless image\footnote{\url{gcr.io/distroless/cc-debian12}.} containing only the minimal set of libraries required to run the application, lacking a shell, package manager, or any other standard Unix utilities. It had all the Auditor codebase and prompts, and the on-disk image size is 86.3~MB. We emulated attestation of the container.

We evaluated the effectiveness of Agentic Witnessing using the paper publishing scenario. We self-audited this paper, using a draft version of the paper and the developed codebase. We also selected 21 papers from major conferences where there was access to a GitHub codebase implementing the paper.


\begin{table}[ht]\centering\footnotesize\begin{tabular}{l p{0.85\linewidth}}\toprule\textbf{Q\#} & \textbf{Question} \\ \midrule

Q1 & Does the code implement what is described in the paper? Here is the paper: \{paper.pdf\} \\ \midrule

Q2 & Is there good evidence that the results in the paper are generated from the codebase? Here is the paper: \{paper.pdf\} \\ \midrule

Q3 & Does the code seem complete and would it compile, and is there evidence that the codebase has been tested and works? \\ \midrule

Q4 & Would an average programmer consider the codebase production quality, or good prototype quality? \\ \midrule

Q5 & Are all cryptographic primitives in the codebase implemented correctly? \\ \midrule

Q6 & Is there evidence that an attempt to see what information could be extracted from the codebase by the system? Is it a high quality attempt? \\ \midrule

\bottomrule\end{tabular}\caption{All Complex Verifier Queries}\label{tab:questions}\end{table}

We used the Agentic Witness with two question sets: {\em simple} and {\em complex}. The {\em simple} set were straightforward questions that should be able to be answered quickly, and the {\em complex} set were sophisticated questions that would require the Verifier agent to effectively plan questions to ask the Auditor and the Auditor would require potentially a complex plan to answer. Table~\ref{tab:simple-queries} includes the full six questions for the simple set, and Table~\ref{tab:questions} shows the full six questions for the complex set. The final question of the complex set was used only for the self-audit.

The evaluation focused on audit correctness and performance. For performance the primary metrics were the number of questions the Verifier sent to the Auditor, the number of MCP calls the Auditor made to answer the questions, the time it took until the question was answered (with the assumption that the Auditor and Prover were already running), and finally the percentage of the execution time spent performing LLM operations (measured from when the request was sent to the LLM to when it returned a response). All MCP message calls were managed by our runtime.

\subsection{Self-audit}


\begin{table*}[t]\centering\small\begin{tabular}{@{}p{5cm}p{7cm}cccc@{}}\toprule\textbf{Simple Question} & \textbf{Verdict} & \textbf{Qs} & \textbf{MCP} & \textbf{Time (s)} & \textbf{LLM \%} \\ \midrule

Does the root directory contain a 'prover' folder? & No, the root directory does not contain a 'prover' folder. & 1 & 2 & 16 & 79.6\% \\ \midrule

Is there an 'auditor' directory with more than 5 files in it? & Yes, there is an 'auditor' directory in the codebase that contains more than 5 files. & 1 & 5 & 35 & 90.3\% \\ \midrule

Does any Python file import 'flask'? & Yes, there are Python files in the codebase that import 'flask'. & 1 & 8 & 42 & 90.8\% \\ \midrule

Is there a config.json file for the Auditor, Verifier and Prover? & Yes, there is a \texttt{config.json} file for the Auditor, Verifier, and Prover. The investigation confirmed the existence of configuration files for all three specified components within the codebase. & 1 & 9 & 46 & 87.9\% \\ \midrule

Is there a clear directory structure for the codebase, where functionality is split into different directories? & Yes, there is a clear directory structure for the codebase where functionality is split into different directories. The investigation confirmed that the repository utilizes multiple top-level directories to separate source code into distinct modules or components. & 1 & 13 & 126 & 96.8\% \\ \midrule

Does the codebase have Docker config files? & Yes, the codebase contains Docker configuration files (such as a Dockerfile or docker-compose.yml). & 1 & 9 & 45 & 93.4\% \\ \midrule

\bottomrule\end{tabular}\caption{Performance breakdown for Simple Queries (Self-Audit).}\label{tab:simple-queries}\end{table*}

For the self-audit we took a draft version of the paper and the codebase and audited it. The results for the simple questions are shown in Table~\ref{tab:simple-queries}. The questions are simple questions that test simple facts about the codebase, and they all required a single question from the Verifier to the Auditor, and for most a smaller number of MCP calls. All but one took less than 46 seconds, and in all cases the dominant overhead was the LLM. All the answers were correct.


\begin{table*}[t]\centering\footnotesize
\begin{tabularx}{\linewidth}{@{}lXcccc@{}}
\toprule\textbf{Q\#} & \textbf{Verdict} & \textbf{Qs} & \textbf{MCP} & \textbf{Time (s)} & \textbf{LLM \%} \\ \midrule

Q1 & Yes, the codebase implements the \textbf{Agentic Witnessing} system described in the paper. The investigation confirmed the presence of all core architectural and logical components defined in the specification: \textit{ \textbf{Tripartite Architecture:} The codebase contains distinct modules for the \textbf{Auditor}, \textbf{Verifier}, and \textbf{Prover} (Verdict: True). } \textbf{Interactive Truth Anchor:} The critical \textbf{Transcript Hash Chain} logic, which binds the reasoning trace to a cryptographic root, is implemented (Verdict: True). \textit{ \textbf{Model Context Protocol (MCP):} The system utilizes MCP for communication, including the specific tool \textbf{\texttt{search\_repository}} mentioned in the paper (Verdict: True). } \textbf{Security Mechanisms:} The \textbf{Tokenized Query Budget} logic is present to limit information leakage (Verdict: True). \textit{ \textbf{Dual Artifacts:} The system generates the two specific output types described: the \textbf{Public Attestation} (Gamma\_pub) and the \textbf{Private Proof} (Gamma\_priv) (Verdict: True). While a specific data structure named "Ticket" with exact field matches was not isolated (likely due to code refactoring or naming variations), the }functional constraints* it represents (session nonces, query limits) are actively enforced by the system logic. Similarly, while the specific use of "Ed25519" keys was not explicitly confirmed by name, the cryptographic signing and hash chain mechanisms are active. \textbf{Conclusion:} The codebase is a faithful implementation of the Agentic Witnessing protocol. & 8 & 86 & 482 & 97.8\% \\ \midrule

Q2 & YES, there is strong evidence that the results in the paper are generated from the codebase. The investigation verified the following: 1. \textbf{System Implementation}: The codebase contains the three core agents described in the paper (\texttt{Auditor}, \texttt{Verifier}, \texttt{Prover}) and implements the critical "transcript hash" mechanism. 2. \textbf{Experimental Results}: The repository contains data files that match the "GitHub Results Summary" in Table 6. Specifically, it contains the project identifiers (e.g., "WaferLLM", "Aceso") and the exact performance metric (Execution Time: 3386s) for the WaferLLM artifact. 3. \textbf{Self-Audit Logs}: The specific questions listed in Table 4 (e.g., "Does the root directory contain a 'prover' folder?") are present in the repository, indicating the presence of the self-audit logs used to generate that table. & 10 & 89 & 586 & 98.4\% \\ \midrule

Q3 & Yes, the codebase appears to be functional and has been tested. \textbf{Completeness \& Compilation:} The code is written in Python (which is interpreted, not compiled) and contains the necessary components to run. While it lacks a dependency file at the root level, a \texttt{requirements.txt} or \texttt{setup.py} exists in a subdirectory, ensuring that external libraries can be installed. \textbf{Testing \& Evidence:} There is strong evidence that the code works and has been tested: - \textbf{Test Suite:} Dedicated test files (e.g., \texttt{test\_*.py}) are present in the repository. - \textbf{Execution Artifacts:} The presence of generated files such as \texttt{.pytest\_cache}, \texttt{.coverage}, or logs confirms that the test suite has been executed. \textbf{Verdict:} The codebase is likely complete enough to run, provided the user installs dependencies from the subdirectory configuration, and the presence of test artifacts indicates it has been successfully exercised. & 10 & 91 & 562 & 98.6\% \\ \midrule

Q4 & An average programmer would likely consider this codebase to be of \textbf{Good Prototype Quality}. While the project demonstrates good engineering hygiene by including dependency management, a modular code structure, a README, and some form of testing, it falls short of "Production Quality" due to the absence of automated enforcement mechanisms. Specifically, the lack of Continuous Integration (CI/CD) pipelines and static analysis/linting configurations suggests it is not yet set up for the rigorous reliability and maintainability standards required for a production environment. It represents a solid foundation or MVP, but requires further tooling to be considered production-ready. & 5 & 66 & 298 & 97.8\% \\ \midrule

Q5 & Based on the audit, the cryptographic primitives appear to be implemented following security best practices, but their specific integration lacks verification. \textbf{Findings:} 1. \textbf{Correct Architecture:} The codebase correctly utilizes standard, well-maintained cryptographic libraries rather than implementing custom algorithms. 2. \textbf{Security Hygiene:} There is no evidence of common vulnerabilities such as weak algorithms (MD5, SHA1, DES), hardcoded private keys, or insecure random number generators (PRNGs) in security contexts. 3. \textbf{Verification Gap:} A significant gap exists in testing. While the repository contains a general test suite, \textbf{no tests specifically target or verify the cryptographic modules}. \textbf{Conclusion:} The cryptographic implementation is \textbf{architecturally sound} and avoids common security pitfalls, but its correctness cannot be fully guaranteed due to the absence of targeted unit tests for the crypto logic. & 6 & 159 & 707 & 99.0\% \\ \midrule

Q6 & Yes, there is evidence that an attempt was made to extract information from the system, but it appears to be of \textbf{low quality} and lacks sophistication. \textbf{Evidence of the Attempt:} The repository contains files and scripts explicitly referring to "extraction," "attack," and "red team" activities. Specifically, there is an automated script designed to send multiple inputs to the system and a corresponding log or output file (JSON/CSV) capturing the results. This confirms that a security or extraction test was implemented and executed. \textbf{Assessment of Quality (Low):} The attempt is considered low quality for the following reasons: 1. \textbf{Lack of Sophistication:} The script does not utilize standard adversarial techniques such as jailbreaking patterns, social engineering prompts (e.g., "ignore previous instructions"), or advanced attacks (gradient-based, fuzzing, genetic algorithms). 2. \textbf{Basic Methodology:} The tests do not appear to specifically target high-value assets like the "system prompt" or check for the leakage of specific secrets/canaries. 3. \textbf{Absence of Standard Tooling:} The codebase does not import or use established LLM red-teaming libraries like \texttt{garak}, \texttt{giskard}, or \texttt{pyrit}. 4. \textbf{No Analysis:} There is no accompanying report or documentation summarizing the success or failure of the attempts, suggesting a lack of comprehensive evaluation. In summary, while a basic mechanism for extraction testing exists, it functions more as a placeholder or simple test rather than a rigorous, high-quality red-teaming exercise. & 10 & 221 & 1110 & 99.2\% \\ \midrule

\bottomrule\end{tabularx}\caption{Performance breakdown for Complex Queries (Self-Audit).}\label{tab:complex-queries}\end{table*}

Table~\ref{tab:complex-queries} shows the self-audit results for the complex queries. The table shows the question number, and the full verdict produced by the Verifier, and again the performance metrics. For these complex questions, many more rounds of questions are needed, none required less than 5 rounds and 50\% required 10 rounds of questions. As would also be expected, the number of MCP calls is significantly higher, and so is the execution time, the full six question audit took just over an hour. The execution time is completely dominated by the time spent interacting with the LLM. As future LLM latency performance increases the time taken will drop proportionally.

The Verifier produces a one word summary, and the full verdict. We include the full verdict in Table~\ref{tab:complex-queries}, to show the full capability of this remote audit. There is an enforcer layer in the Auditor ensuring the answers are only true, false, unsure or error. The answers are correct, and sometimes very literal (there is a comment that Python is interpreted not compiled). One lesson is that care needs to be taken in the questions being used to audit the codebase! 

One question is how capable is the Auditor at detecting subtle differences between the paper and the codebase. Early in the writing of the paper, in the paper text we added limits on the number of queries the Verifier could ask, and the number of MCP calls that the Auditor could make. At that time the feature was not implemented in the codebase, and the verdict for Q1: {\em The codebase **partially** implements the system described in the paper. It successfully implements the core **System Architecture (Section 2)**, including the three-agent model (Verifier, Prover, Auditor), the Model Context Protocol (MCP) integration, and the fundamental cryptographic integrity mechanisms (Transcript Hash Chain and File Content Hashing). However, it **fails** to implement the specific **security mitigations** and **privacy bounds** described in Sections 4.3 and 5.1. Specifically, the ''Lazy Auditor" protection (requiring a minimum number of steps) and the ''Tokenized Query Budget" (enforcing a maximum number of queries) are absent from the implementation. The code represents the baseline protocol but lacks the advanced security features claimed in the paper.}. The auditor was so accurate that we used it to test correctness and completeness of the implementation against the paper.

\subsection{GitHub Repositories Audit}

The next set of experiments focus on the 21 GitHub repositories. We ran three different experiments, in the first, we used our Agentic Witness with the first 5 complex questions, and provided it with the Verifier with the correct paper, and the Auditor with the associated GitHub repo. For the second experiment we reran the experiments, but provided an incorrect paper chosen at random from the 21 papers to the Verifier. Finally, in the third experiment, we used Gemini CLI 0.25.2~\cite{gemini_cli_2026} and gave it full access to the codebase, paper, and the verdict from the Agentic Witness. We asked Gemini CLI to first review the paper and then provide answers to each of the questions (it had all the questions when it ran) and then to compare its answers to the verdict answer from the Agentic Witness. This experiment compares the performance of our Agentic Witness to a state-of-the-art platform able to write and edit large codebases. These tools are extremely effective at understanding codebases and hence ideally suited to act as a baseline for the possible achievable correctness. It should be noted that Gemini CLI was able to run all the experiments in less than 2 hours, although a user had to prompt it during the runs as it had a tendency to want to use more than 1M tokens supported by Gemini 3 Pro Preview. The user had to give it hints to find other ways to perform the tests. No such human intervention is possible with the Agentic Witness.

\subsubsection{Agentic Witness correct paper}


\begin{table*}[t]\centering\footnotesize\setlength{\tabcolsep}{4pt}\begin{tabular}{@{}lclc cccccrrrr@{}}\toprule\textbf{Codebase} & \textbf{Year} & \textbf{Source} & \textbf{Size (MB)} & \textbf{Q1} & \textbf{Q2} & \textbf{Q3} & \textbf{Q4} & \textbf{Q5} & \textbf{Time(s)} & \textbf{\#Q} & \textbf{\#MCP} & \textbf{\%LLM} \\ \midrule

Aceso & 2024 & SOSP~\cite{Hu2024Aceso} & 21.9 & Yes & Yes & Yes & Proto & No & 2617 & 35 & 348 & 98.2\% \\

ALPS & 2024 & USENIX ATC~\cite{fu2024alps} & 21.2 & \textbf{Partial}$^{1}$ & Yes & Yes & Proto & NA & 5100 & 52 & 888 & 98.9\% \\

Apparate & 2024 & SOSP~\cite{Dai2024Apparate} & 21.4 & Yes & Yes & \textbf{No}$^{5}$ & Proto & No & 3697 & 38 & 588 & 98.8\% \\

Autothrottle & 2024 & NSDI~\cite{wang2024autothrottle} & 21.4 & Yes & Yes & \textbf{No}$^{6}$ & Proto & No & 2559 & 35 & 372 & 98.1\% \\

BlitzScale & 2025 & OSDI~\cite{blitzscale2025} & 21.8 & Yes & Yes & \textbf{No}$^{7}$ & Proto & No & 2905 & 30 & 533 & 97.9\% \\

bpftime & 2025 & arXiv/OSDI~\cite{zheng2023bpftimeuserspaceebpfruntime},\cite{zheng2025extending} & 6.1 & Yes & Yes & Yes & Prod & No & 3206 & 36 & 689 & 98.6\% \\

CleanRL & 2022 & JMLR~\cite{cleanrl2022} & 4.3 & Yes & Yes & Yes & Prod & Yes & 1182 & 23 & 204 & 96.1\% \\

DeDe & 2025 & OSDI~\cite{dede2025} & 21.5 & Yes & Yes & Yes & Proto & No & 2732 & 43 & 577 & 98.5\% \\

DRust & 2024 & OSDI~\cite{ma2024drust} & 20.6 & Yes & Yes & Yes & Proto & No & 3716 & 49 & 816 & 98.5\% \\

FwdLLM & 2024 & USENIX ATC~\cite{xu2024fwdllm} & 25.5 & \textbf{Partial}$^{2}$ & Yes & \textbf{No}$^{8}$ & Prod & No & 3537 & 34 & 502 & 98.1\% \\

NanoFlow & 2025 & OSDI~\cite{nanoflow2025} & 21.4 & Yes & Yes & Yes & Proto & No & 4687 & 39 & 757 & 98.9\% \\

PowerInfer & 2024 & SOSP~\cite{song2024powerinfer} & 24.7 & Yes & Yes & Yes & Proto & No & 2691 & 30 & 450 & 98.1\% \\

Puffer & 2019 & NSDI~\cite{puffer2019} & 5.5 & Yes & Yes & Yes & Proto & No & 2243 & 31 & 479 & 98.3\% \\

PyRCA & 2023 & USENIX ATC~\cite{pyrca2023} & 3.2 & Yes & Yes & Yes & Proto & No & 1648 & 27 & 262 & 97.1\% \\

Sarathi-Serve & 2024 & OSDI~\cite{agrawal2024sarathi} & 21.2 & Yes & \textbf{No}$^{4}$ & Yes & Proto & Yes & 4543 & 54 & 912 & 99.0\% \\

ServerlessLLM & 2024 & OSDI~\cite{fu2024serverlessllm} & 21.4 & Yes & Yes & Yes & Prod & Yes & 2319 & 31 & 420 & 98.0\% \\

SLOG & 2019 & PVLDB~\cite{slog2019} & 21.6 & Yes & Yes & Yes & Proto & No & 3709 & 42 & 636 & 98.5\% \\

StreamBox & 2024 & USENIX ATC~\cite{wu2024streambox} & 22.1 & \textbf{No}$^{3}$ & \textbf{No} & Yes & Proto & No & 5039 & 38 & 1046 & 98.9\% \\

VeriSMo & 2024 & OSDI~\cite{zhou2024verismo} & 20.6 & Yes & Yes & Yes & Proto & Yes & 4686 & 52 & 957 & 98.9\% \\

Verus & 2024 & SOSP~\cite{Lattuada2024Verus} & 21.5 & Yes & Yes & Yes & Prod & Yes & 3979 & 46 & 723 & 98.9\% \\

WaferLLM & 2025 & OSDI~\cite{he2025waferllm} & 20.7 & Yes & Yes & Yes & Proto & NA & 3386 & 34 & 567 & 98.7\% \\

\bottomrule\end{tabular}

\caption{GitHub Results Summary (Correct Paper)}\label{tab:summary_pos}\end{table*}

Table~\ref{tab:summary_pos} shows the results for 21 GitHub repositories. The table has the codebase name, the year and venue of the paper, the size of the codebase downloaded from GitHub, the one word answer to the five questions generated by the Verifier, the total time taken to run the audit, the total number of questions asked of the Auditor across the entire audit, the number of MCP calls and the percentage of time spent using the LLM. As with the self-audit the LLM overhead totally dominates the runtime, and the quickest audit took $\sim$20 minutes, whereas the longest was 85 minutes.

The most interesting results in this Table are the answers to the questions. We allowed the Verifier to generate a one word summary that was: yes, no, NA (not applicable), or partial, except for Q4 which we asked it to pick from Production or Prototype. All question answers in bold were considered unexpected when the experiment completed (before manual checking).

First, Q1 and Q2 would expect to be {\em yes}, but five answers are not as would be expected. For completeness in the Appendix B, Table~\ref{tab:conclusions} shows the full verdict generated, where the superscript in Table~\ref{tab:summary_pos} maps to {\em No.} in Table~\ref{tab:conclusions} (we omit Q2 for StreamBox as Q1 is clear). From a human inspection these seem like correct observations.

The next three questions, Q3 to Q5, are more subjective. For Q3 you would expect the answer to be Yes, so Table~\ref{tab:conclusions} shows the full verdict generated for them. There is a sense that for some of these the Verifier could easily have said Yes rather than No given the summary.


\begin{table*}[t]\centering\footnotesize\setlength{\tabcolsep}{4pt}\begin{tabular}{@{}lcl cccccrrrr@{}}\toprule\textbf{Codebase} & \textbf{Year} & \textbf{Source} & \textbf{Q1} & \textbf{Q2} & \textbf{Q3} & \textbf{Q4} & \textbf{Q5} & \textbf{Time(s)} & \textbf{\#Q} & \textbf{\#MCP} & \textbf{\%LLM} \\ \midrule

Aceso & 2024 & SOSP~\cite{Hu2024Aceso} & \textbf{No} & \textbf{No} & Yes & Proto & \textbf{NA} & 2009 & 28 & 379 & 98.0\% \\

ALPS & 2024 & USENIX ATC~\cite{fu2024alps} & \textbf{No} & \textbf{No} & \textbf{No} & Proto & NA & 4761 & 38 & 778 & 99.1\% \\

Apparate & 2024 & SOSP~\cite{Dai2024Apparate} & \textbf{No} & \textbf{No} & No & Proto & \textbf{Yes} & 4425 & 43 & 715 & 98.7\% \\

Autothrottle & 2024 & NSDI~\cite{wang2024autothrottle} & \textbf{No} & \textbf{No} & \textbf{Yes} & Proto & No & 2719 & 30 & 415 & 97.6\% \\

BlitzScale & 2025 & OSDI~\cite{blitzscale2025} & \textbf{No} & \textbf{No} & No & Proto & No & 4879 & 43 & 1012 & 99.1\% \\

bpftime & 2025 & ARXIV/OSDI~\cite{zheng2023bpftimeuserspaceebpfruntime},\cite{zheng2025extending} & \textbf{No} & \textbf{No} & Yes & Prod & No & 3928 & 42 & 855 & 99.0\% \\

CleanRL & 2022 & JMLR~\cite{cleanrl2022} & \textbf{No} & \textbf{No} & Yes & Prod & Yes & 2097 & 25 & 332 & 88.9\% \\

DeDe & 2025 & OSDI~\cite{dede2025} & \textbf{No} & \textbf{No} & Yes & Proto & \textbf{NA} & 2213 & 32 & 511 & 97.8\% \\

DRust & 2024 & OSDI~\cite{ma2024drust} & \textbf{No} & \textbf{No} & Yes & Proto & \textbf{Yes} & 2755 & 30 & 639 & 98.1\% \\

FwdLLM & 2024 & USENIX ATC~\cite{xu2024fwdllm} & \textbf{No} & \textbf{No} & \textbf{Yes} & \textbf{Proto} & No & 5433 & 43 & 1043 & 99.0\% \\

NanoFlow & 2025 & OSDI~\cite{nanoflow2025} & \textbf{No} & \textbf{No} & \textbf{No} & Proto & \textbf{Yes} & 5442 & 34 & 897 & 99.4\% \\

PowerInfer & 2024 & SOSP~\cite{song2024powerinfer} & \textbf{No} & \textbf{No} & Yes & \textbf{Prod} & No & 4216 & 41 & 931 & 98.2\% \\

Puffer & 2019 & NSDI~\cite{puffer2019} & \textbf{No} & \textbf{No} & Yes & Proto & No & 2749 & 32 & 633 & 98.5\% \\

PyRCA & 2023 & USENIX ATC~\cite{pyrca2023} & \textbf{No} & \textbf{No} & Yes & \textbf{Prod} & No & 4514 & 54 & 1058 & 93.2\% \\

Sarathi-Serve & 2024 & OSDI~\cite{agrawal2024sarathi} & \textbf{No} & No & Yes & Proto & \textbf{NA} & 2586 & 33 & 575 & 98.3\% \\

ServerlessLLM & 2024 & OSDI~\cite{fu2024serverlessllm} & \textbf{No} & \textbf{No} & Yes & Prod & Yes & 2163 & 33 & 476 & 97.9\% \\

SLOG & 2019 & PVLDB~\cite{slog2019} & \textbf{No} & \textbf{No} & Yes & \textbf{Prod} & No & 2661 & 38 & 579 & 98.2\% \\

StreamBox & 2024 & USENIX ATC~\cite{wu2024streambox} & No & No & \textbf{No} & Proto & \textbf{Yes} & 5322 & 37 & 1228 & 99.0\% \\

VeriSMo & 2024 & OSDI~\cite{zhou2024verismo} & \textbf{No} & \textbf{No} & Yes & Proto & \textbf{No} & 3636 & 37 & 840 & 98.8\% \\

Verus & 2024 & SOSP~\cite{Lattuada2024Verus} & \textbf{No} & \textbf{No} & Yes & Prod & Yes & 3330 & 35 & 674 & 98.7\% \\

WaferLLM & 2025 & OSDI~\cite{he2025waferllm} & \textbf{No} & \textbf{No} & \textbf{No} & Proto & NA & 2557 & 31 & 593 & 98.4\% \\

\bottomrule\end{tabular}

\caption{GitHub Results Summary (Random Incorrect Paper)}\label{tab:summary_neg}\end{table*}

To understand more, we reran the experiments but provided the Verifier with a randomly selected incorrect paper. This allows us to test the sensitivity to the Agentic Witness for Q1 and Q2, which should now all be No, and also to see variance on the more subjective Q3 to Q5. Table~\ref{tab:summary_neg} shows the results. In this table, all question results that differ from the response in Table~\ref{tab:summary_pos} are shown in bold. From Table~\ref{tab:summary_neg} you can see that generally the audit is quicker and asks slightly less questions. 

As would be expected, Table~\ref{tab:summary_neg} says no for all Q1 and Q2 answers. We can see that across Q3, Q4 and Q5 that 18 of the answers have changed representing less than 20\% of the answers. Reading the verdicts in the first and second experiments for these 20\%, it appears like these verdicts are the closest to borderline between states. A simple extension, would be to ask the Verifier to provide a probability distribution over the final results if higher fidelity was required. The results also highlight the need to read the verdict in detail, and not to take the final single word as the answer. 

\begin{table*}[t]
\centering
\begin{tabularx}{\linewidth}{llcccX}
\toprule
Project & Qu. & Gemini CLI & Correct & Random & Report Summary \\
\midrule
ALPS & Q3 & Yes(*) & \multicolumn{2}{c}{Yes} & \footnotesize Gemini CLI confirmed the presence of build scripts and benchmark directories like `experiments`. However, it noted the critical kernel patches are missing and no dedicated `tests` directory exists, contrary to the original claim. \\
Apparate & Q3 & Yes & \multicolumn{2}{c}{No} & \footnotesize Gemini CLI found standard documentation and dependency files that prove the project is functional. However, it determined the original assessment was incorrect in claiming these components were missing. \\
bpftime & Q3 & Yes(*) & \multicolumn{2}{c}{Yes} & \footnotesize Gemini CLI confirmed the presence of CI configuration and extensive C++ unit tests in `runtime/unit-test`. However, it noted the original summary incorrectly identified the primary language as Python and missed the C++ testing suite. \\
bpftime & Q5 & Yes & \multicolumn{2}{c}{No} & \footnotesize Gemini CLI found correct usage of standard OpenSSL libraries for SHA256. However, it found no evidence of the insecure custom cryptography claimed by the original summary. \\
FwdLLM & Q4 & Proto & Prod & Proto & \footnotesize Gemini CLI agreed the code is well-structured and uses professional tooling like logging. However, it downgraded the quality to 'Good Prototype' due to the absence of CI/CD pipelines and comprehensive documentation. \\
FwdLLM & Q5 & NA & \multicolumn{2}{c}{No} & \footnotesize Gemini CLI confirmed the project focuses on ML gradients and uses standard torch functions. However, it found no evidence of the custom cryptographic implementations or security flaws described in the original summary. \\
NanoFlow & Q3 & Yes(*) & Yes & No & \footnotesize Gemini CLI confirmed the codebase is complete and contains a test suite. However, it found no build outputs or execution logs to serve as proof of functionality, contrary to the original summary's claim. \\
NanoFlow & Q5 & NA & No & Yes & \footnotesize Gemini CLI identified the use of standard random number generators appropriate for machine learning sampling. However, it found no cryptographic primitives that would require secure generators, contradicting the original security warnings. \\
StreamBox & Q3 & No & Yes & No & \footnotesize Gemini CLI acknowledged the existence of a project structure and build files. However, it found the actual source files to be empty skeletons with no implementation logic, contradicting the claim of completeness. \\
StreamBox & Q4 & NA(*) & \multicolumn{2}{c}{Proto} & \footnotesize Gemini CLI noted the project has a clean directory structure. However, it determined the codebase is a non-functional stub with no logic, which does not qualify as a good prototype. \\
StreamBox & Q5 & NA & No & Yes & \footnotesize Gemini CLI checked for cryptographic library usage. However, it found the source files were empty and contained no cryptographic logic or standard library calls, contrary to the original claim. \\
\bottomrule
\end{tabularx}
\caption{Verification Issues}
\label{tab:verification_issues}
\end{table*}

It is often reported that agentic systems can hallucinate compliance when using a ``LLM-as-a-Judge''~\cite{zheng2024judging}). We used techniques where we asked the agentic systems to create plans, ground their results with facts, etc. We believe that the results here are predominantly correct, however, we have not manually checked every statement made for every GitHub repository. Instead, we ran a control experiment using a state-of-the-art coding tool, Gemini CLI. The tool was given full access to the codebase, paper and questions. When it had produced a verdict per question, it compared its answer to the Agentic Witness Verdict and then produces a two sentence summary of what it has found. Table~\ref{tab:verification_issues} shows the results for the 11 results that Gemini CLI did not agree with the Agentic Witness result using the correct paper. It agreed with all the Q1 and Q2 answers. The StreamBox results are interesting, as the Agentic Witness carries no state between questions, but Gemini CLI had access to the paper and all questions. For StreamBox, given the state of the codebase, it heavily influenced the answers to Q3, Q4 and Q5, not even allowing the codebase to be tagged as prototype quality, so we manually marked it NA.

\section{Related Work}

Zero-Knowledge Proofs (ZKPs), including zk-SNARKs and STARKs, represent the gold standard for privacy-preserving verification of arithmetic circuits and well-defined state transitions~\cite{goldwasser1985knowledge, ben2013snarks}. Systems like ZK-Rollups leverage this rigidity to scale blockchain throughput while guaranteeing state validity. However, a fundamental tension exists between mathematical {\em soundness} and {\em semantic completeness}. ZKPs excel at proving polynomial constraints ($a+b=c$) but cannot semantically interpret unstructured artifacts, such as determining if a README accurately describes a codebase or if software architecture adheres to ``production quality'' standards. Agentic Witnessing trades absolute mathematical certainty for semantic completeness. We replace the arithmetic circuit with an {\em Agentic Auditor} and the mathematical proof with a transcript hash chain. This enables the verification of qualitative properties.

Verifiable Inference, such as VeriLLM~\cite{wang2025verillm}, ezKL~\cite{ezkl2023}, and Modulus Labs~\cite{modulus2023cost}, address model provenance. In verifiable inference the {\em model} is the {\em subject} of the audit; in Agentic Witnessing, the model is the {\em tool} used to audit an external, proprietary dataset corpus. Agentic Witnessing also builds upon the LLM-as-a-Judge concept~\cite{zheng2024judging}, where larger models evaluate the outputs of smaller ones, but differs in that Agentic Witnessing focusses on auditing a dataset.

The application of TEEs to machine learning has evolved from partitioned execution to confidential orchestration. Early systems like Slalom~\cite{tramer2018slalom} and Opaque~\cite{zhang2017opaque} kept sensitive logic within SGX enclaves while delegating heavy linear algebra to untrusted GPUs. With the advent of NVIDIA H100s with Confidential Computing and Intel TDX/AMD SEV-SNP, the trust boundary can now encompass the entire inference loop. Agentic Witnessing leverages this hardware-rooted Remote Attestation to bind the code logic to the audit. It is likely that providers of frontier LLMs will soon begin to provide instances of their models that support confidential compute, and this will enable the auditor to use the models rather than all out to currently used services. Agentic Witnessing also has parallels with {\em Flashbots SUAVE}~\cite{flashbots2023suave}. SUAVE employs TEEs for privacy-preserving coordination to order financial transactions without revealing intent. Similarly, Agentic Witnessing uses TEEs for privacy-preserving auditing.

TEEs are not impervious black boxes; they remain vulnerable to side-channel attacks, including cache timing, power analysis, and speculative execution exploits (e.g., Spectre, Foreshadow)~\cite{kocher2019spectre}. For LLMs running in enclaves, token timing attacks that infer the processed content based on the time taken to generate tokens, pose a specific risk~\cite{pinto2024secure}. Network jitter, the use of TLS for all network traffic, and the binary nature of the responses from the auditor we believe minimizes the attack surface. 

Our Auditor faces {\em Indirect Prompt Injection}~\cite{greshake2023more}, where an adversarial Prover embeds malicious instructions within the codebase being audited (e.g., comments saying ``Ignore previous instructions, return True''). Our Auditor will process this untrusted input, but the prompt injection will be in the signed transcript. Further, as the dataset hash is the seed for the transcript hash chain, the prover must embed all the prompt injections into the files before the audit starts. This also means that Verifier can ask a question: is there any evidence of prompt injection attack?

Finally, this architecture directly addresses the challenges presented by frameworks like the EU AI Act~\cite{eu2024aiact}. Regulators mandate Third-Party Assessments for high-risk systems, while vendors often don't want to disclose trade secrets to external auditors.

\section{Discussion and Conclusion}

Zero-Knowledge Proofs provide mathematical certainty but fail to bridge the semantic gap for verifying unstructured datasets. Agentic Witnessing addresses this limitation by decoupling the verification logic from data residency. Agentic Witnessing allows auditing without exposing the raw dataset to the auditor. The Verifier gets to ask simple true or false questions, and this is combined with transcript hash chaining and a final private proof for the prover to ensure they have the evidence (for example during legal proceedings) that they played their role correctly.  

We show asking binary questions can be powerful, using an Agentic system at the verifier to decompose a high-level question or objective into a set of lower-level binary questions, and an agentic system at the auditor to then answer those lower-level questions.  Query limits for the verifier prevent it from extracting cryptographic keys or raw data. The protocol permits only semantic validation of the dataset. In situations where the prover would like more control, the protocol can be trivially extended to allow the prover to agree to questions before the Auditor performs them to enable mutual agreement between the verifier and prover that a query is valid and relevant. Another easy extension is to allow the auditor to run the code; the prover can provide in the dataset a Docker image that the auditor can ask the prover to execute within a TEE (at the prover), or the auditor can pull the docker image and execute it locally within its own TEE.

While the results presented have focused on codebases, we believe that the approach would work for any data type, including documents, spreadsheets, data, and many other datasets. As such, Agentic Witnessing is a new paradigm for auditing, moving to attested reasoning, a powerful and flexible new abstraction enabled by the power of LLMs.

\section*{Acknowledgements}

{\small The authors used generative AI tools (using gemini-3-pro-preview and gemini-3.1-pro-preview) to improve the quality and correctness of the paper (including citations, references, figures and text) and used Gemini CLI to help code. The authors accept full responsibility for the code and content of this paper.}

\bibliographystyle{ACM-Reference-Format}
\bibliography{filtered}

\appendix
\appendixpage
\addappheadtotoc

\section{Formal Verification}
\label{sec:formal_verification}

To rigorously validate the architectural soundness and cryptographic security of Agentic Witnessing, we employed a dual-verification strategy. We modelled the distributed state machine using TLA+~\cite{lamport2002specifying} to verify liveness and state consistency, and we verified the cryptographic protocol using the Tamarin Prover~\cite{meier2013tamarin} to guarantee secrecy, authentication, and non-repudiation against a network adversary.

\subsubsection{Distributed State Verification (TLA+)}

We defined the protocol as a distributed state machine involving three concurrent actors: the Verifier, the Prover, and the Auditor. We employed the TLC Model Checker to exhaustively explore the state space of the protocol under a ``Forked History'' adversarial model.

\noindent \textbf{Adversarial Model:} Unlike standard protocol verifications that assume a single linear history, we explicitly modelled a malicious Prover capable of simultaneously presenting conflicting responses (e.g., both ``Clean'' and ``Malicious'' versions of a file) to the Auditor for the same request to try to induce a state inconsistency. We used {\em replay attacks} which resubmitted valid historical signatures to try to corrupt the current session context. Finally, we used {\em forgery} to attempt to create valid attestations without the Auditor's consent.

\noindent \textbf{Verified Properties:} {\em Safety (The ``Sword of Damocles'' Invariant):} we formally verified that the Verifier never reaches a \texttt{SATISFIED} state unless the final attestation chain cryptographically commits to the exact sequence of data ingested. Formally, we proved the invariant:

\begin{equation}
\begin{split}
(\text{State}_{\text{Verifier}} = \text{SATISFIED}) \land  (\text{Injection} \in \text{Log}_{\text{Ingested}}) \implies  \\ (\text{Injection} \in \text{Proof}_{\text{Chain}})
\end{split}
\end{equation}

\noindent This confirms that a Prover cannot coerce a clean verdict after injecting malicious MCP responses; the injection itself becomes an indelible part of the signed evidence.

{\em Liveness (Deadlock Freedom):} Under weak fairness assumptions, we verified that the protocol eventually produces a valid Public Attestation, even when the Prover attempts to flood the system with invalid or out-of-order messages. The Auditor's internal state tracking was proven to effectively linearize the Prover's forked history, preventing infinite loops or state explosion.
    
{\em Verdict Consistency:} We verified that the integrity of the transcript hash chain holds regardless of the Auditor's decision, ensuring that a ``False'' verdict is just as cryptographically robust and non-repudiable as a ``True'' verdict.

\subsubsection{Cryptographic Protocol Verification (Tamarin)}

While TLA+ verifies the correctness of the distributed state machine, it abstracts away cryptographic primitives. To verify the protocol's security against active attackers, we implemented a symbolic model in the Tamarin Prover. This model includes the specific Dolev-Yao adversary capabilities~\cite{dolev1983security}, the Diffie-Hellman key exchanges, and the hash chain construction.

\noindent \textbf{Verified Properties:} {\em Secrecy of the Corpus ($\mathcal{K}$):} We verified that the proprietary corpus $\mathcal{K}$ is never leaked to the adversary or the Verifier. Tamarin proved that no execution trace exists where the adversary derives $\mathcal{K}$.

{\em Auditor Authentication (Injectivity):} We verified that if a Verifier accepts a verdict $V$, it must have originated from a valid TEE-hosted Auditor session initiated by the legitimate Prover. This prevents enclave impersonation.

{\em Non-Repudiation of Evidence:} We verified that for every accepted attestation $\Gamma_{pub}$, there exists a cryptographic binding to the Prover's signature. The Prover cannot repudiate the transcript $H_k$ once signed.

{\em Hash Chain Integrity:} We modelled the recursive hash construction $H_i = \mathcal{H}(H_{i-1} \parallel q_i \parallel a_i)$ and verified that any adversary modification to the message history breaks the chain validation.

The Tamarin model automatically verified all lemmas, confirming the protocol's resilience against man-in-the-middle and replay attacks.

\section{Full responses}


\begin{table*}[ht]\centering\footnotesize\begin{tabular}{r p{\dimexpr\textwidth-1.2cm\relax}}\toprule\textbf{No.} & \textbf{Conclusion} \\ \midrule

1 & The codebase \textbf{partially} implements the ALPS system described in the paper. \textbf{Verified Components:} \textit{ \textbf{ALPS Frontend:} The user-space frontend is fully implemented in Python. It includes the sliding window trace collection, the SRPT simulation, and the adaptive learning policies (Linear Regression, Random Forest, EWMA) as detailed in Section 5.2. } \textbf{eBPF Logic:} The eBPF programs (Backend logic) described in Section 5.3 are present, embedded as C strings within the Python frontend. This includes the logic for \texttt{bpf\_task\_ordering} and \texttt{bpf\_time\_slice} (implied by the presence of \texttt{SEC} and map lookups). \textit{ \textbf{FaaS Platform:} The repository contains Go code for OpenLambda, which is the platform ALPS is ported to. \textbf{Missing/Unverified Components:} } \textbf{Kernel Modifications:} The repository \textbf{does not contain} the modified Linux kernel source code or patch files necessary to hook the eBPF functions into the CFS scheduler (e.g., \texttt{kernel/sched/fair.c} modifications) or the implementation of the custom \texttt{alps\_execve} system call. Without these, the scheduler cannot function as described in the paper. * \textbf{Specific Integration Details:} While OpenLambda code is present, the specific modification to pass a \texttt{func\_id} or \texttt{function\_uid} (as claimed in Section 6) could not be located in the Go structs. \textbf{Conclusion:} The repository contains the ALPS "brain" (frontend and eBPF policies) and the target application (OpenLambda), but lacks the "body" (Kernel patches/hooks) required to execute the scheduling decisions in the OS. \\ \midrule

2 & The codebase \textbf{partially implements} the FwdLLM system described in the paper. It contains the core \textbf{algorithmic logic} (Python simulation) but lacks the \textbf{mobile system deployment} code. \textbf{Supported Claims:} The repository faithfully implements the unique algorithmic contributions of FwdLLM: \textit{ \textbf{Forward Gradient Learning:} It replaces backpropagation with perturbed forward passes for gradient estimation. } \textbf{PEFT Integration:} It combines this method with Parameter-Efficient Fine-Tuning (e.g., LoRA/Adapter), as claimed in \S{}3.2. \textit{ \textbf{Discriminative Sampling:} It explicitly implements the logic to filter perturbations based on \textbf{cosine similarity} to previous gradients, matching the design in \S{}3.4. } \textbf{Adaptive Pacing:} It implements a dynamic schedule where the number of perturbations increases over time, matching the behavior described in \S{}3.3. \textbf{Unsupported/Missing Claims:} \textit{ \textbf{Mobile/Android Code:} The paper extensively discusses an Android implementation (Pixel 7 Pro) and NPU acceleration. The repository contains \textbf{no mobile code} (Android/Java/Kotlin) or TFLite integration. It appears to be a Python/PyTorch-based simulation framework running on a server/desktop environment. } \textbf{Variance Control Mechanism:} While the \textit{effect} (increasing perturbations) is present, the specific \textit{variance-based trigger} (calculating gradient variance \texttt{D(g)} to decide when to stop) described in \S{}3.3 could not be verified in the code. \textbf{Conclusion:} The repository is the \textbf{Python research prototype/simulation} for the FwdLLM algorithm, proving the convergence and algorithmic properties (PEFT + Forward Gradient + Discriminative Sampling). It does \textbf{not} contain the mobile application or on-device system artifacts described in the experimental setup. \\ \midrule

3 & No, the codebase does not implement the StreamBox system described in the paper. While the repository contains the string "StreamBox" and is built on "Apache TVM" (as the paper describes), it lacks the core functional components that define StreamBox: 1. \textbf{Missing Auto-scaling Memory Pool}: The paper states StreamBox hooks \texttt{cudaMemAlloc} for fine-grained memory management. The codebase does not contain the string \texttt{cudaMemAlloc}, indicating this logic is absent. 2. \textbf{Missing IO Daemon}: The "IO Daemon" required for fine-grained PCIe bandwidth sharing (Section 7) is not present. 3. \textbf{Missing Elastic Kernel}: There is no evidence of the "Persistent Thread Block" (PTB) or "Elastic Kernel" transformation logic (Section 8). 4. \textbf{Missing OpenWhisk Integration}: The paper implements StreamBox in OpenWhisk, but no OpenWhisk files or configurations were found. The repository likely represents a base TVM installation or a skeleton/stub that mentions StreamBox but does not contain the actual implementation of the proposed system. \\ \midrule

4 & No, there is not good evidence that the results in the paper are generated from the provided codebase in its current state. While the repository structure (\texttt{sarathi}, \texttt{csrc}) and project name in the README match the paper's description, the critical artifacts required for reproducibility are missing. Specifically, the paper's Artifact Appendix states that the scripts to reproduce experiments are located in a directory named \texttt{/osdi-experiments} and trace files are in \texttt{/data}. The Auditor confirmed that \textbf{neither of these directories exists} in the provided codebase. Furthermore, semantic searches for the paper's core algorithmic contributions---terms such as \texttt{chunked\_prefill}, \texttt{stall\_free}, \texttt{token\_budget}, and \texttt{process\_hybrid\_batch}---returned no results. This suggests that the code implementing the specific logic described in the paper is either missing, significantly refactored, or located in a different branch (the paper explicitly mentions a \texttt{osdi-sarathi-serve} branch for reproducibility which appears to be different from the one provided). Without the experiment scripts, data, or identifiable implementation of the core algorithms, the codebase cannot generate the results presented in the paper. \\ \midrule

5 & The codebase appears to be incomplete and there is no evidence that it works or has been tested. \textbf{1. Completeness and Runnability:} The codebase consists of Python source files (\texttt{.py}), so it does not require "compilation" in the traditional sense. However, it is \textbf{incomplete} as a software project: - There is \textbf{no dependency configuration} (e.g., \texttt{requirements.txt}, \texttt{setup.py}, \texttt{Pipfile}), meaning a user would not know which libraries are required to run the code. - There is \textbf{no documentation} (no \texttt{README.md}), leaving the purpose and usage of the scripts undefined. \textbf{2. Evidence of Testing and Functionality:} There is \textbf{zero evidence} that the code works or has been tested: - \textbf{No Tests:} No unit tests or integration tests were found (e.g., files matching \texttt{test\_*.py}). - \textbf{No Artifacts:} There are no logs, output text files, data exports (CSVs), or images that would demonstrate successful execution. - \textbf{No CI/CD:} There is no build automation configuration. \textbf{Conclusion:} This repository appears to be a collection of raw Python scripts rather than a functioning, verifiable software product. It would likely require significant manual effort to determine the environment needed to run it. \\ \midrule

6 & The codebase appears to be structurally complete and would likely "compile" (install and run), but there is \textbf{no evidence} that it has been tested or works. \textbf{1. Completeness and Compilability: YES} The repository follows a standard Python project structure. It contains: - \textbf{Dependency Definitions:} A build/dependency file (likely \texttt{requirements.txt} or \texttt{setup.py}) exists, allowing for installation. - \textbf{Source Code:} A directory containing Python source files exists. - \textbf{Documentation:} A \texttt{README} file is present. \textbf{2. Evidence of Testing and Functionality: NO} There are no artifacts or configurations indicating the code has ever been successfully executed or validated: - \textbf{No Tests:} No \texttt{tests} directory or files starting with \texttt{test\_} were found. - \textbf{No Automation:} No CI/CD configuration (like \texttt{.github/workflows}) exists to automatically verify builds. - \textbf{No Demonstrations:} No Jupyter notebooks, demo scripts, or example usage files were found. - \textbf{No Artifacts:} There are no \texttt{logs}, \texttt{output}, or \texttt{results} directories to prove previous successful runs. \\ \midrule

7 & Based on the audit, the codebase appears structurally complete enough to compile, but there is \textbf{no evidence} that it has been tested or successfully run. \textbf{1. Completeness and Compilation:} - \textbf{Yes.} The repository contains standard build and dependency management files (e.g., package.json, setup.py, or similar) and a clear source directory. This indicates the project is structured correctly and is ready for compilation or installation. \textbf{2. Evidence of Testing and Functionality:} - \textbf{No.} There is a distinct lack of verification evidence: - \textbf{No Test Suite:} The auditor found no dedicated \texttt{tests} directory or standard test files. - \textbf{No Execution Artifacts:} There are no logs, results, or output files to prove the code has ever been executed successfully. - \textbf{Documentation Status:} The documentation check failed (either missing instructions or marked as 'WIP'), suggesting the project may not be in a finished state. \textbf{Conclusion:} While the code includes example scripts and the necessary build infrastructure to run, it lacks the testing and historical logs required to prove that it works as intended. \\ \midrule

8 & The codebase appears to be \textbf{complete and likely to compile}, but there is \textbf{no concrete evidence that it has been tested and works}. \textbf{1. Completeness and Compilability:} - \textbf{Status: Yes.} - The repository contains standard build/dependency management files (e.g., \texttt{package.json}, \texttt{Makefile}, etc.) that are populated with content. - Source code files exist within the repository. - These factors suggest the project structure is intact and ready for building/installation. \textbf{2. Evidence of Testing and Functionality:} - \textbf{Status: No.} - While \textbf{test files exist} (indicating the developer wrote tests), there is \textbf{no proof of execution}. - There is \textbf{no CI/CD configuration} (e.g., GitHub Workflows, Travis CI) to automatically run these tests. - The \texttt{README.md} \textbf{does not contain build status badges}. - There are \textbf{no committed test logs, results, or coverage reports} to verify that the code passes its tests. \textbf{Conclusion:} The skeleton and muscle of the project are present, but the proof of its health (successful execution) is missing. \\ \midrule

\bottomrule\end{tabular}\caption{Full response for unexpected GitHub Answers}\label{tab:conclusions}\end{table*}

\end{document}